\documentclass[%
 aip,
 jcp,
 twocolumn,
 amsmath,amssymb,
 eprint=false,
reprint,%
]{revtex4-1}

\usepackage{graphicx}% Include figure files
\usepackage{dcolumn}% Align table columns on decimal point
\usepackage{bm}% bold math
\usepackage[utf8]{inputenc}
\usepackage[T1]{fontenc}
\usepackage{mathptmx}
\usepackage{etoolbox}
\usepackage{siunitx}
\usepackage{booktabs}
\usepackage{epstopdf}
\usepackage{array,multirow}
\usepackage[hidelinks]{hyperref}

\makeatletter
\def\@email#1#2{%
 \endgroup
 \patchcmd{\titleblock@produce}
  {\frontmatter@RRAPformat}
  {\frontmatter@RRAPformat{\produce@RRAP{*#1\href{mailto:#2}{#2}}}\frontmatter@RRAPformat}
  {}{}
}%
\makeatother

\begin{document}

\preprint{AIP/123-QED}

%%%%%%%%%%%%%%%%%%%%%%%%%%%%%%%%%%%%%%%%%%%%%%%%%%%%%%%%%%%%%%%%%%%%%
%% The document title should be given as usual. Some journals require
%% a running title from the author: this should be supplied as an
%% optional argument to \title.
%%%%%%%%%%%%%%%%%%%%%%%%%%%%%%%%%%%%%%%%%%%%%%%%%%%%%%%%%%%%%%%%%%%%%
\title{\Large Excited State Properties from the Bethe--Salpeter Equation: \\
State-to-State Transitions and Spin-Orbit Coupling}

\author{Paula Himmelsbach}
\affiliation
{Institute of Theoretical Solid State Physics, Karlsruhe Institute of Technology, Kaiserstra\ss e 12, 76131 Karlsruhe, Germany}

\author{Christof Holzer$^*$}
\email[Email for correspondence: ]{holzer@kit.edu}
\affiliation
{Institute of Theoretical Solid State Physics, Karlsruhe Institute of Technology, Kaiserstra\ss e 12, 76131 Karlsruhe, Germany}

\date{\today}% It is always \today, today,

%%%%%%%%%%%%%%%%%%%%%%%%%%%%%%%%%%%%%%%%%%%%%%%%%%%%%%%%%%%%%%%%%%%%%
%% Some journals require a list of abbreviations or keywords to be
%% supplied. These should be set up here, and will be printed after
%% the title and author information, if needed.
%%%%%%%%%%%%%%%%%%%%%%%%%%%%%%%%%%%%%%%%%%%%%%%%%%%%%%%%%%%%%%%%%%%%%
%\keywords{natural virtual orbitals, random phase approximation, GW approximation, vertex correction}

%% Each author should be given as a separate \author command.
%%
%% Corresponding authors should have an e-mail given after the author
%% name as an \email command. Phone and fax numbers can be given
%% using \phone and \fax, respectively; this information is optional.
%%
%% The affiliation of authors is given after the authors; each
%% \affiliation command applies to all preceding authors not already
%% assigned an affiliation.
%%
%% The affiliation takes an option argument for the short name.  This
%% will typically be something like "University of Somewhere".
%%
%% The \altaffiliation macro should be used for new address, etc.
%% On the other hand, \alsoaffiliation is used on a per author basis
%% when authors are associated with multiple institutions.
%%%%%%%%%%%%%%%%%%%%%%%%%%%%%%%%%%%%%%%%%%%%%%%%%%%%%%%%%%%%%%%%%%%%%

\begin{center}
\date{\today}% It is always \today, today,
\end{center}

\begin{abstract}
The formalism to calculate excited state properties from the
$GW$--Bethe--Salpeter equation (BSE) method is introduced, 
providing convenient access to excited state absorption,
excited state circular dichroism, and excited state optical 
rotation in the framework of the $GW$--BSE method.
This is achieved using the second-order transition density, 
which can be obtained by solving a set of auxiliary equations 
similar to time-dependent density functional theory (TD-DFT). 
The proposed formulation therefore leads to no increase in the 
formal computational complexity when compared to the 
corresponding ground state properties.
We further outline the calculation of fully relaxed
spin-orbit coupling matrix elements within the $GW$--BSE method, 
allowing us to include perturbative corrections for 
spin-orbit coupling in aforementioned properties. 
These corrections are also extended to TD-DFT. 
Excited state absorption and perturbative spin-orbit 
coupling corrections within $GW$--BSE are evaluated for 
a selected set of molecular systems, yielding promising results.
\end{abstract}

%\section{TOC Graphic}

%\begin{figure}[htbp]
%\includegraphics{TOC.eps}
%\caption{For Table of Contents Only}
%\label{fig:TOC}
%\end{figure}

\maketitle

\section{Introduction}

Recently, analytic implementations of excited state properties from the 
$GW$--Bethe--Salpeter equation (BSE) method have surged in literature.
For example, excited state dipole moments,
\cite{Knysh.Villalobos-Castro.ea:Excess.2023}
non-linear optical properties as hyperpolarizabilities 
and two-photon absorption have become available in the framework
of the $GW$--BSE method\cite{Rauwolf.Klopper.ea:Non-linear.2024}
and approximate Lagrangian Z-vectors methods related to
excited state geometry optimizations.
\cite{Villalobos-Castro.Knysh.ea:Lagrangian-Z-vector.2023}
Going beyond the simple calculation of excited state energies 
and oscillator strengths is an important task in the development 
of $GW$--BSE, leading into one of the final domains where 
time-dependent density functional
theory (TD-DFT) is more versatile than $GW$--BSE due to the lack 
of available implementations of the latter.
This situation is unsatisfactory given the excellent
performance of $GW$--BSE for both local and 
charge-transfer excited states.
\cite{Blase.Attaccalite:Charge-transfer.2011,Gui.Holzer.ea:Accuracy.2018,
Blase.Duchemin.ea:Bethe–Salpeter.2020,Cho.Bintrim.ea:Simplified.2022,
Li.Golze.ea:Combining.2022,Yao.Golze.ea:All-Electron.2022}
Recent advanced within the $GW$--BSE method further outline
the possibility to perform frequency-dependent 
simulations,\cite{Loos.Blase:Dynamical-correction.2020,Bintrim.Berkelbach:Full-frequency.2022,Loos.Romaniello:Static-and-dynamic.2022} 
and other ways to recover double excitations.
\cite{Monino.Loos:Spin-Conserved.2021}

In this work, we will therefore introduce the formalism required to 
perform calculations on excited state properties within the framework
of the $GW$--BSE method. 
Examples of the accessible properties include
excited state absorption, excited state circular dichroism, 
and excited state optical rotation. We will focus on the
static screened BSE,\cite{Blase.Duchemin.ea:Bethe–Salpeter.2020}
and note that an extension to dynamic BSE methods
\cite{Loos.Blase:Dynamical-correction.2020,
Bintrim.Berkelbach:Full-frequency.2022,
Loos.Romaniello:Static-and-dynamic.2022} 
is in principle possible, but not straightforward.
Given the high similarity of the required intermediates, 
we will also outline a way to perturbatively
include spin-orbit coupling (SOC) in these properties in both the 
ground and excited state. This allows for a convenient evaluation 
of these properties at the $GW$--BSE level of theory, 
further encouraging this powerful method to prosper in 
fields that have been out of reach before. 
We will additionally use the opportunity
to introduce spin-orbit coupling matrix elements (SOCMEs) in a less
incorrect way than previous TD-DFT implementations,
\cite{Zhendong-Li.Liu:Combining.2013,
Gao.Bai.ea:Evaluation.2017,
Souza.Farias.ea:Predicting.2019,
Liao.Kasper.ea:State.2023}
and put them more in line with previous works on multiconfigurational
linear response functions.\cite{Vahtras.ea:Spin–orbit-coupling.1992}
Combining the possibility to access excited state properties and
spin-orbit coupling (SOC), we will finally outline the capabilities 
and performance of the newly developed method for four selected sizable
molecular systems. Favorable aspects as well as caveats of the 
$GW$--BSE method for excited state properties and perturbative 
spin-orbit coupling for these systems will subsequently be discussed. 

\section{Theory}

\subsection{State-to-state transitions in the framework of the $GW$-Bethe--Salpeter equation method}
\label{subsec:bse_linear}

Similar to TD-DFT, excitation energies can be extracted from the poles
of the symplectic BSE eigenvalue problem
\cite{PhysRevLett.80.3320,
Faber.Boulanger.ea:Many-body.2013,
Blase.Duchemin.ea:BetheSalpeter.2018,
Krause.Klopper:Implementation.2017,
Holzer.Klopper:Ionized.2019} 
\begin{align}
\label{eq:quadratic}
\left[ (\mathbf{A} - \mathbf{B}) (\mathbf{A} + \mathbf{B}) \right] \mathbf{R}^{N} = & \omega_{N}^2 \mathbf{R}^{N} \, \text{,} \\
\left[ (\mathbf{A} + \mathbf{B}) (\mathbf{A} - \mathbf{B}) \right] \mathbf{L}^{N} = & \omega_{N}^2 \mathbf{L}^{N}\, \text{.}
\end{align}
Within the static screened BSE, the matrices $\mathbf{(A+B)}$ and $\mathbf{(A-B)}$ are defined as
\begin{align}
\label{eq:apb}
(A+B)_{{i}a,{j}b} = & (\epsilon_a - \epsilon_{{i}}) \delta_{ab} \delta_{{i}{j}} + 
H^{+}_{ai,bj} \text{,} \\
\label{eq:amb}
(A-B)_{{i}a,{j}b} = & H^{-}_{ai,bj} \text{.}
\end{align}
The solutions $\mathbf{R}=(\mathbf{X}+\mathbf{Y})$ and 
$\mathbf{L}=(\mathbf{X}-\mathbf{Y})$ are normalized to 
obey the relation
\begin{equation}
\langle R^N | L^M \rangle = \delta_{NM} \text{.}
\end{equation}
This partition is useful, as after a unitary transformation
into the real atomic orbital (AO) basis, the matrices $\mathbf{(A+B)}$ 
and $\mathbf{R}$ are symmetric, while $\mathbf{(A-B)}$ 
and $\mathbf{L}$ are skew-symmetric.
\cite{Bauernschmitt.Ahlrichs:Treatment.1996,Bauernschmitt.Ahlrichs:Stability.1996}
Eqs.~(\ref{eq:apb}) and (\ref{eq:amb}) have introduced the 
two-electron kernel contributions,
\begin{align}
\label{eq:bse_kernel}
    H^{+,\text{BSE}}_{pq,rs} =&\ v_{pq,rs} - W_{ps,qr}(\omega=0) - W_{pr,qs}(\omega=0) \text{,}\\
    H^{-,\text{BSE}}_{pq,rs} =&\ W_{ps,qr}(\omega=0) -  W_{pr,qs}(\omega=0) \text{,}
\end{align}
with $v_{pq,rs}$ being a Coulomb integral, and $\mathbf{W}(\omega)$ 
denoting the screened exchange. The latter is obtained from 
$\mathbf{v}$ and the inverse of the dielectric function $\mathbf{\kappa}$
\begin{equation}
\label{eq:W_BSE}
    W_{pq,rs}(\omega) = \sum_{tu} \kappa_{pq, tu}^{-1}(\omega) v_{tu,rs}\,\text{.}
\end{equation}
Efficient procedures to evaluate Eq.~(\ref{eq:W_BSE}) 
for arbitrary values of $\omega$ have been outlined in literature.\cite{Holzer.Klopper:Ionized.2019,Holzer:Practical.2023}
To emphasize the similarity between the BSE and TD-DFT, we
note that the latter simply replaces the kernels with
\begin{align}
    H^{+,\text{DFT}}_{pq,rs} =&\ v_{pq,rs} + f^{\text{XC, sy}}_{ps,qr} \text{,}\\
    H^{-,\text{DFT}}_{pq,rs} =&\ f^{\text{XC, as}}_{pr,qs} \text{.}
\end{align}

Similar to DFT\cite{Parker.Rappoport.ea:Quadratic.2018}
and the calculation of two-photon absorption 
and hyperpolarizabilities
in the framework of the BSE,
\cite{Rauwolf.Klopper.ea:Non-linear.2024} a state-to-state
transition within the BSE can be calculated from the double
residue of the quadratic response function
\begin{equation}
\label{eq:hyp_trace}
\begin{split}
     v_{NM} = & \lim_{\omega_{\eta} \rightarrow -\Omega_N}  (\omega_{\eta} + \Omega_N) \lim_{\omega_{\theta} \rightarrow \Omega_M} (\omega_{\theta} - \Omega_M) \\
  \times  & \langle \langle v^{\zeta}; v^{\eta} (\omega_{\eta}), v^{\theta}(\omega_{\theta}) \rangle \rangle  
  =   Tr\left( \hat{v}^{\zeta} \gamma^{NM} \right) 
\end{split}
\end{equation}
where $\gamma^{NM}$ is the state-to-state transition density matrix. 
As it constitutes a single determinant method,
$\gamma^{NM}$ can be obtained from the idempotency constraint of
the time-dependent Kohn--Sham determinant as
\cite{Furche:On.2001}
\begin{align}
\label{eq:idem0}
    \gamma^{0} = & \gamma^{0} \gamma^{0} \\
    \gamma^{N} = & \gamma^{0} \gamma^{N} + \gamma^{N} \gamma^{0} \\
    \label{eq:gamma_2}
    \gamma^{NM} = & \gamma^{0} \gamma^{NM} + \gamma^{N} \gamma^{M} + \gamma^{M} \gamma^{N} + \gamma^{NM} \gamma^{0} \text{.}
\end{align}
Given the independence of the density matrix from the
actual eigenvalues, Eq.~(\ref{eq:idem0})
still holds true for $GW$ methods that only affect the eigenvalues.
This is at least true for the common $G_0W_0$ method, but
also for the eigenvalue self-consistent $GW$ variant.
\cite{PhysRevLett.96.226402,
Kotani.Van-Schilfgaarde.ea:Quasiparticle.2007}
Inserting the definitions of $\gamma^{0}$ and $\gamma^{N}$, 
\begin{align}
    \gamma^{0}(x, x') = & \sum_{j} \phi_j(x) \phi_j^{*}(x') \\
    \gamma^{N}(x, x') = & \sum_{bj} X_{bj}^N \phi_j(x) \phi_b^{*}(x') + Y_{jb}^N \phi_b(x) \phi_j^{*}(x')\text{,}
\end{align}
as determined from the ground state Kohn--Sham equations ($\phi_j$) and 
linear response ($\{ X, Y \}$) leads to the matrix form
\begin{equation}
\label{eq:twoparticle}
    \gamma^{N M} =
    \begin{pmatrix}
        K^{N M}_{ij} & X^{N M}_{ib} \\
        Y^{N M}_{aj} & K^{N M}_{ab}
    \end{pmatrix}
    \text{,}
\end{equation}
of the second-order transition density $\gamma^{N M}$.
Eq.~(\ref{eq:twoparticle}) closely resembles the corresponding 
quantities in Refs.~\citenum{Rauwolf.Klopper.ea:Non-linear.2024} 
and \citenum{Parker.Rappoport.ea:Quadratic.2018}. Note that 
$\gamma^{N M}$ differs from the second-order 
single-particle reduced BSE density matrix of 
Ref.~\citenum{Rauwolf.Klopper.ea:Non-linear.2024} by
replacing the perturbations $\{ \eta, \zeta\}$ with the excited states
$\{N, M \}$. Notably, the matrix elements are calculated slightly
differently, too. As indicated in Eq.~(\ref{eq:hyp_trace}), 
a switch of sign has taken place for the excitation.
\cite{Parker.Rappoport.ea:Quadratic.2018} 
Instead of the eigenpair $\omega_M, \{\mathbf{X}$, $\mathbf{Y} \}$, 
the time-reversal symmetry related eigenpair 
$-\omega_M, \{\mathbf{Y}^{*}$, $\mathbf{X}^{*} \}$ 
is to be used. For purely real orbitals, this manifests in
a change of sign of $\mathbf{L}^M$. This change of sign must be accounted
for in subsequent evaluations of the matrix elements of $\gamma^{N M}$.
The diagonal subblocks of Eq.~(\ref{eq:twoparticle}) are obtained 
in a straightforward manner as products of linear response 
vectors as
\begin{align}
    K^{N M,+}_{ij} =&\ - {\frac{1}{4}}\sum_a \left[ R_{ia}^{N} R_{ja}^{M} + L_{ia}^{M} L_{ja}^{N} + (N \leftrightarrow M) \right] \,\text{,}\\
    K^{N M,-}_{ij} =&\ \phantom{-} {\frac{1}{4}}\sum_a \left[ L_{ia}^{N} R_{ja}^{M} + R_{ia}^{N} L_{ja}^{M} - (N \leftrightarrow M) \right] \,\text{,}\\
    K^{N M,+}_{ab} =&\ \phantom{-} {\frac{1}{4}}\sum_i \left[ R_{ia}^{N} R_{ib}^{M} + L_{ia}^{N} L_{ib}^{M} + (N \leftrightarrow M) \right] \,\text{,}\\
    K^{N M,-}_{ab} =&\ \phantom{-} {\frac{1}{4}}\sum_i \left[ L_{ia}^{N} R_{ib}^{M} + R_{ia}^{N} L_{ib}^{M} - (N \leftrightarrow M) \right] \,\text{,}
\end{align}
arising from the second and third term on the 
right hand side of Eq.~(\ref{eq:gamma_2}).
$\mathbf{R}$ and $\mathbf{L}$ are solutions of the linear response 
equations for the excited states $N$ and $M$, were
the time-reversal symmetric partner of the excitation $M$ is to be used. 

The off-diagonal blocks $X^{N M}_{ib}$ and $Y^{N M}_{aj}$ 
require the solution of the second-order Bethe--Salpeter response 
equations. Again using the symmetry adapted linear combinations
$\mathbf{R}^{NM}=\left(\mathbf{X}^{NM} + \mathbf{Y}^{NM}\right)$ and 
$\mathbf{L}^{NM}=\left(\mathbf{X}^{NM} - \mathbf{Y}^{NM}\right)$, 
these equations read
\begin{align}
\label{eq:quadratic_sy}
\left[ (\mathbf{A} - \mathbf{B}) (\mathbf{A} + \mathbf{B}) - \omega_{\zeta}^2 \mathbf{1} \right] \mathbf{R}^{N M} = & \mathbf{U}^{N M}\, \text{,} \\
\label{eq:quadratic_as}
\left[ (\mathbf{A} + \mathbf{B}) (\mathbf{A} - \mathbf{B}) - \omega_{\zeta}^2 \mathbf{1} \right] \mathbf{L}^{N M} = & \mathbf{V}^{N M}\, \text{.}
\end{align}
The right-hand side (RHS) of Eqs.~(\ref{eq:quadratic_sy})
and (\ref{eq:quadratic_as}) are defined as 
\begin{equation}
\label{eq:uhyper}
\begin{split}
    U^{N M}_{ia} = & -\frac{1}{2} \sum_j \left[R^{M}_{ja} M^{N}_{ij} + L^{M}_{ja} N^{N}_{ij} + (N \leftrightarrow M) \right] \\
    & + \frac{1}{2} \sum_b \left[R^{M}_{ib} M^{N}_{ab} + L^{M}_{ib} N^{N}_{ab} + (N \leftrightarrow M) \right] \\
    & + H^{+}_{ia} [K^{N M, +}_{kl}] + H^{+}_{ia} [K^{N M, +}_{cd}] + g^{sy}_{ia} (R^{N}, R^{M}) \,,
\end{split}
\end{equation}
\begin{equation}
\label{eq:vhyper}
\begin{split}
    V^{N M}_{ia} = & +\frac{1}{2} \sum_j \left[L^{M}_{ja} M^{N}_{ij} + R^{M}_{ja} N^{N}_{ij} - (N \leftrightarrow M) \right] \\
    & - \frac{1}{2} \sum_b \left[L^{M}_{ib} M^{N}_{ab} + R^{M}_{ib} N^{N}_{ab} - (N \leftrightarrow M) \right] \\
    & - H^{-}_{ia} [K^{N M, -}_{kl}] - H^{-}_{ia} [K^{N M, -}_{cd}] 
    -g^{as}_{ia}(L^{N}, L^{M}) \,\text{.}
\end{split}
\end{equation}

In Eqs.~(\ref{eq:uhyper}) and (\ref{eq:vhyper}), the quantities
$\mathbf{M}$ and $\mathbf{N}$ are further defined as
\begin{align}
    M^{N}_{pq} =& \sum_{rs}  H^{+}_{pq,rs} R_{rs}^N  \,,\\
    N^{N}_{pq} =& \sum_{rs}  H^{-}_{pq,rs} L_{rs}^N \text{.}
\end{align}
$g^{XC}$ refers to the first hyperkernel. 
Within adiabatic TD-DFT, it is approximated as
\begin{equation}
\label{eq:dft_hyperkernel}
  g^{XC}(x, x', x'') \approx \frac{\partial^3 E (\rho)}{ \partial \rho (x)  \partial \rho (x') \partial \rho (x'')} \text{.}
\end{equation}
The hyperkernel in Eq.~(\ref{eq:dft_hyperkernel}) is symmetric
for the local density (LDA) and generalized gradient 
approximation (GGA). Starting from metaGGAs, antisymmetric
terms from the current density arise, 
leading to non-vanishing antisymmetric hyperkernel contributions in 
Eq.~(\ref{eq:vhyper}). \cite{Bates.Furche:Harnessing.2012,
Holzer.Franzke.ea:Assessing.2021} This holds true
in cases of current-free\cite{Bates.Furche:Harnessing.2012} 
and current-carrying ground states.\cite{Pausch.Holzer:Linear.2022,Holzer.Franzke.ea:Current.2022}
Spin integration further leads to the hyperkernel vanishing in 
cases were the states $N$ and $M$ have different spin symmetries.
Within the BSE, the hyperkernel is approximated in a similar manner
as the linear response kernel, for which the approximation
\begin{equation}
    \frac{\partial G{W}}{ \partial G } \approx {W}
\end{equation}
is used.
\cite{Blase.Duchemin.ea:BetheSalpeter.2018,Villalobos-Castro.Knysh.ea:Lagrangian-Z-vector.2023} 
Accordingly, we assume 
\begin{equation}
    g^{\text{BSE}} = \frac{\partial^2 G{W}}{ \partial G \partial G} \approx \frac{\partial {W}}{\partial G} \approx 0 \text{,}
\end{equation}
to be equally valid, resulting in no significant hyperkernel 
contribution from the BSE. This approximation has been shown 
to lead to negligible errors for the Lagrangian Z-vector equations, 
and we project this also being the case for excited state properties.
\cite{Villalobos-Castro.Knysh.ea:Lagrangian-Z-vector.2023}
Note that for hybrid TDDFT/BSE methods, 
as for example the correlation-kernel
augmented BSE,\cite{Holzer.Klopper:Communication.2018}
both the DFT and (vanishing) BSE hyperkernel are formally
required. An implementation of the correlation-kernel
augmented BSE is however trivial once both non-linear
implementations are available. 
The frequency of the electric field perturbation $\omega^{\zeta}$ is 
formally set to $(\Omega_N - \Omega_M)$ when solving the
second-order Bethe--Salpeter response equations (\ref{eq:quadratic_sy}) 
and (\ref{eq:quadratic_as}). This leads to a set of equations for each
pair of states. Special precaution needs to be taken in case of 
an actual excited state being in the vicinity of this energy difference. 
In this case, adding a non-vanishing imaginary component 
can be used to guarantee numerical stability.
\cite{Rauwolf.Klopper.ea:Non-linear.2024}
To partly reduce computational complexity and also resolve issues
with the pole structure, it has been suggested to 
simply set $\omega^{\zeta} = 0$ in a pseudo-wavefunction approach.
\cite{Alguire.Ou.ea:Calculating.2015,Ou.Alguire.ea:Derivative.2015}
This is a valid approximation for differences
$|\Omega_N - \Omega_M| << \Omega_1 $, i.e. if the first excited state
has a significantly higher energy than the energy difference between 
the excited states $N$ and $M$.

By inserting Eqs.~(\ref{eq:twoparticle}) and (\ref{eq:vhyper}) 
into Eq.~(\ref{eq:hyp_trace}), a state-to-state coupling 
matrix element of a symmetric operator can be evaluated as
\begin{equation}
\label{eq:sy_trace}
\begin{split}
 v^{\zeta, sy}_{NM} = & Tr \left( v^{\zeta}_{ij} K^{NM,+}_{ij} \right) 
 + Tr \left( v^{\zeta}_{ab} K^{NM,+}_{ab} \right) \\
 + & Tr \left( v^{\zeta}_{ia} R^{NM}_{ia} \right) \text{.}
\end{split}
\end{equation} 
The most prominent example of a symmetric operator $\hat{v}$ is 
the dipole operator,
\begin{equation}
    v_{pq}^{\zeta,\text{dip.}} = \langle \phi_p | \vec{r}_{\zeta} | \phi_q \rangle \text{,}
\end{equation}
which, when used for states of the same multiplicity, will 
lead to the excited state transition dipole moment components.

\subsection{Spin-orbit coupling matrix elements}
\label{subsec:SOCMEs}

The coupling matrix elements for skew-symmetric operators
can be obtained from the skew-symmetric part
\begin{equation}
\label{eq:as_trace}
\begin{split}
v^{\zeta, as}_{NM} = &  Tr \left( v_{ij}^{\zeta} K^{NM,-}_{ij} \right) + Tr \left( v_{ab}^{\zeta} K^{NM,-}_{ab} \right) \\
+ & Tr \left( v_{ia}^{\zeta} L^{NM}_{ia} \right) \text{.}
\end{split}
\end{equation}
The most important example in this work is 
the spin-orbit mean field operator (SOMF) 
\cite{Hess.Marian.ea:mean-field.1996,
Helmich-Paris.Hattig.ea:Spin-Free.2016}
\begin{equation}
\begin{split}
    v_{pq}^{\zeta,\text{SOMF}} = &  \left( \phi_p \left| h_1^{\zeta} \right|  \phi_q \right) + 2 \sum_{r=1}^{\text{occ.}} \left\{ \left( \phi_r \phi_r \left| h_2^{\zeta}  \right| \phi_p \phi_q \right) \right. \\
    - & \left. \frac{3}{2} \left( \phi_p \phi_r \left| h_2^{\zeta}  \right| \phi_r \phi_q \right)
    + \frac{3}{2} \left( \phi_r \phi_q \left| h_2^{\zeta}  \right| \phi_p \phi_r \right) \right\}
\end{split}
\end{equation}
with
\begin{align}
    h^{\eta} _1 = & \frac{1}{2c} \sum_K \frac{Z_K \left[ \vec{r}_K \times \vec{p} \right]_{\eta}}{\vec{r}_K^3} \\
    h^{\eta} _2 = & \frac{1}{2c} \frac{\left[ \vec{r} \times \vec{p} \right]_{\eta}}{\vec{r}^3} \text{,}
\end{align}
leading to the well known spin-orbit coupling matrix elements (SOCMEs).
The final term in Eq.~(\ref{eq:sy_trace}) is not present in 
Eq.~(\ref{eq:as_trace}), as any skew-symmetric operator
is traceless.

To arrive at previous results, 
\cite{Gao.Bai.ea:Evaluation.2017,
Souza.Farias.ea:Predicting.2019,
Liao.Kasper.ea:State.2023}
two approximations must be made in Eq.~(\ref{eq:as_trace}). 
First, the final term, yielding the
orbital response is neglected in previous works. Second, 
the Tamm-Dancoff approximation needs to be applied
also for full TD-DFT vectors, leading to the observed
re-normalization necessary in these previous works.
SOCMEs calculated from Eq.~(\ref{eq:as_trace}) can therefore
be regarded as being more consistent
from a formal point of view. This correctness however
comes at the cost of solving the auxiliary second-order
response equations, i.e. Eqs.~(\ref{eq:quadratic_sy}) and
(\ref{eq:quadratic_as}), as well as a more elaborate 
evaluation needed for the diagonal terms of $\gamma^{NM}$.
The resulting equations are in fact similar to those
obtained in the early works of Vahtras \textit{et al},
\cite{Vahtras.ea:Spin–orbit-coupling.1992} safe for the
DFT or BSE specific kernels and hyperkernels.

We further note that the first-order matrix element
\begin{equation}
 v_{0N}^{\eta, as} = Tr \left( v_{ia} L^N_{ia} \right) \text{,}
\end{equation}
is correctly evaluated using the $\mathbf{L} = (\mathbf{X}-\mathbf{Y})$ part instead of the 
$\mathbf{R} = (\mathbf{X}+\mathbf{Y})$ part as suggested in 
Refs.~\citenum{Gao.Bai.ea:Evaluation.2017}, \citenum{Souza.Farias.ea:Predicting.2019}, and 
\citenum{Liao.Kasper.ea:State.2023}. 
We stress that after a unitary transformation
to the AO basis set, $(\mathbf{X}+\mathbf{Y})$ is symmetric, while 
the SOMF operator is skew-symmetric. Therefore,
$Tr (v_{\mu \nu}^{\text{SOMF}} (X+Y)_{\mu \nu})$ 
actually vanishes for every excited state. Only
within the Tamm-Dancoff approximation, where $\mathbf{Y}=0$, 
this issue is rectified.

\subsection{State Coupling using Perturbative Spin-Orbit Coupling}

The state interaction between singlet and triplet 
excited states can be obtained from quasi-degenerate
perturbation theory (QDPT) by the means of an effective 
Hamiltonian\cite{Souza.Farias.ea:Predicting.2019}
\begin{equation}
\label{eq:qdpt}
    \hat{H}_{\text{rel}} = \hat{H}^0 + \hat{H}^{\text{SOC}} \text{.}
\end{equation}

The zeroth-order Hamiltonian $\hat{H}_0$ is a diagonal matrix
\begin{equation}
    \hat{H}^0 = 
    \begin{pmatrix}
        0 &  & & \\
          & \omega_N & &  \\
          &  & \omega_M &  \\
          & & & \dots
    \end{pmatrix}\text{,}
\end{equation}
with zero representing the ground state, followed by the
singlet and triplet excitation energies. We
note that the $m_L = \{\pm 1, 0\}$ triplet excitation 
energies are still energetically degenerate in $\hat{H}_0$. The 
matrix elements of the spin-orbit coupling Hamiltonian 
are subsequently obtained from $v^{\bar{\eta}, \text{SOMF}}_{NM}$,
\begin{equation}
    \hat{H}^{\text{SOC}}_{NM} = \mathbf{S} \langle N | \hat{H}^{SO}_{\eta} | M \rangle = \mathbf{S} v^{{\eta}, \text{SOMF}}_{NM} \text{,}
\end{equation}
after a transformation from the Cartesian $\eta \in \{x,y,z\}$ 
into the spherical basis $\bar{\eta} \in \{\pm 1, 0\}$ basis using the 
unitary transformation matrix
\cite{Fernandez-Corbaton.Beutel.ea:Computation.2020}
\begin{equation}
    \mathbf{S} = 
    \begin{pmatrix}
    \frac{1}{\sqrt{2}} & \frac{i}{\sqrt{2}} & 0 \\    
    \frac{-1}{\sqrt{2}} & \frac{i}{\sqrt{2}} & 0 \\  
    0 & 0 & 1 \\
    \end{pmatrix}\text{.}
\end{equation}
The selection rules for non-vanishing matrix elements are 
described in detail elsewhere.
\cite{Souza.Farias.ea:Predicting.2019, Liao.Kasper.ea:State.2023}

After the diagonalization of the Hamiltonian $\hat{H}_{\text{rel}}$, 
the perturbed excited state energies $\widetilde{\omega}$ are obtained 
directly from the eigenvalues.
Perturbed transition properties are subsequently 
evaluated similar to Eq.~20 of
Ref.~\citenum{Souza.Farias.ea:Predicting.2019}.
Using the perturbed eigenvector coefficients, $C_N^M$ and $D_{M,m_L}^N$, assigned to singlet and triplet parts 
of the eigenvectors obtained from $\hat{H}_{\text{rel}}$ 
and $m_L=\{ -1, 0, 1\}$, an explicit formula
for perturbed transition properties can be obtained by
projecting the unperturbed transition properties using the
perturbed eigenvectors as
\begin{equation}
\label{eq:dipmom}
\widetilde{v}_{IJ} =
 \underbrace{\sum_{N=0}^{N_s} \sum_{M=0}^{N_s}  C_N^I C_M^{J} v_{NM}^{s-s}}_{\text{singlet part}} \\
  + \underbrace{\sum_{m_L}^{\pm 1,0} \sum_{N=1}^{N_t} \sum_{M=1}^{N_t} D_{N,m_L}^I D_{M, m_L}^{J}
  v_{NM}^{t-t}}_{\text{triplet part}} \text{.}
\end{equation}
Eq.~(\ref{eq:dipmom}) differs from previous results
\cite{Souza.Farias.ea:Predicting.2019} only
by realizing that the fully relaxed singlet-singlet 
and triplet-triplet transition properties 
$v_{NM}$ obtained from Eqs.~(\ref{eq:sy_trace}) 
or (\ref{eq:as_trace}) are to be used. 

If $I=0$, only matrix elements
$v_{0N}^{s-s}$ are needed for the first term in 
Eq.~(\ref{eq:dipmom}), as other contributions
vanish due to the block structure of the effective
Hamiltonian.

\section{Perturbed excited state properties}

Further properties can be evaluated from the corrected transition moments.
To obtain SOC perturbed transition properties, the unperturbed transition property
must be calculated for each singlet-singlet and triplet-triplet transition 
used in the construction of the effective SOC Hamiltonian, 
and subsequently projected using Eq.~(\ref{eq:dipmom}).
The main computational bottleneck are however the second-order Bethe–Salpeter 
response equations (\ref{eq:quadratic_sy}) and (\ref{eq:quadratic_as}), 
which are independent of the actual transition property.
Evaluating the expectation values over different
transition properties is therefore negligible as is the 
projection from the unperturbed to the perturbed representation.

\subsection{Dipole transitions between excited states}

The oscillator strength between two excited states can be obtained as
\begin{align}
    {f}_{IJ} = & \frac{2\Delta \widetilde{\Omega}_{IJ}}{3} |\widetilde{{\mu}}_{IJ}^{l}| \\
    {f}_{IJ} = & \frac{2}{3\Delta \widetilde{\Omega}_{IJ}} |\widetilde{{\mu}}_{IJ}^{v}| \text{,}
\end{align}
from the electric transition dipole moment $\widetilde{{\mu}}_{IJ}$ in either its length
or velocity representation. Further, $\Delta \widetilde{\Omega}_{IJ}= \widetilde{\omega}_I - \widetilde{\omega}_J$
denotes the difference between the excited states.
These improved transition moments provide an excellent starting point for 
excited state dynamics in systems with significant spin-orbit coupling.
Interconversion rates between the vibrational 
ground states of two excited states, $k^{r, 0-0}_{I \rightarrow J}$, 
are subsequently obtained as\cite{Haneder.Da-Como.ea:Controlling.2008,
Kuhn.Weigend:Phosphorescence-lifetimes.2014}
\begin{equation}
 k^{r, 0-0}_{I \rightarrow J} = \frac{2\pi e^2}{\epsilon_0 c^3 m_e h^2} {\Delta \widetilde{\Omega}_{IJ}^2 {f}_{IJ}} \text{.}
\end{equation}

\subsection{Rotatory strength between excited states}

The rotatory strength between two excited states can be calculated as
\begin{equation}
    R_{IJ}^{\alpha \beta} = \text{Im} \left[ \widetilde{{\mu}}_{IJ}^{\alpha} \widetilde{{m}}_{IJ}^{\beta} \right] \text{,}
\end{equation}
$\alpha, \beta$ refer to the respective Cartesian components, $\widetilde{{\mu}}_{IJ}$ 
being the perturbed electric transition dipole moment in either its length or velocity representation, 
and $\widetilde{{m}}_{IJ}$ denoting the perturbed magnetic transition dipole moment.
Only the latter is fully gauge invariant.\cite{Warnke.Furche:Circular.2012}
The rotatory strength is directly related to circular dichroism (CD) and 
optical rotation (OR) spectra. Both properties can be calculated from it 
using standard procedures.\cite{Warnke.Furche:Circular.2012}
The same holds true for the orientation-independent
hermitian optical rotation tensor,\cite{Buckingham.Dunn:Optical.1971}
\begin{equation}
    B_{IJ}^{\alpha \beta} = \left[ R_{IJ}^{\alpha \beta} + R_{IJ}^{\beta \alpha} + \frac{1}{3} A_{IJ}^{\alpha \beta} \right] \text{,}
\end{equation}
where the electric dipole–electric quadrupole polarizability tensor $A$
is\cite{Autschbach:Time-Dependent.2011}
\begin{equation}
    A_{IJ}^{\alpha \beta} = \epsilon_{\alpha \gamma \delta} {\widetilde{\mu}}_{IJ}^{\gamma} {\widetilde{\Theta}}_{IJ}^{\delta \beta}
      + \epsilon_{\beta \gamma \delta} {\widetilde{\mu}}_{IJ}^{\beta} {\widetilde{\Theta}}_{IJ}^{\delta \alpha} \text{.}
\end{equation}
$\widetilde{\Theta}_{IJ}$ denotes the perturbed electric quadrupole transition moment and 
$\epsilon_{\alpha \beta \gamma}$ is the Levi--Civita symbol.

\subsection{Approximate perturbed T-matrices for excited states}

Transition matrix (T-matrix) techniques have recently become rather popular in \textit{ab initio}
modeling of optical properties of materials starting from atomistic simulations. Those can be
obtained in an approximate manner from a sum-over-states expansion as
\cite{Fernandez-Corbaton.Beutel.ea:Computation.2020}
\begin{equation}
\label{eq:tmatrix}
\begin{split}
    T^{m_l, n_l}_I(\omega) = & \frac{i c_h Z_h k_h^3}{6 \pi} \sum_{\alpha, \beta} S_{m_l}^{\alpha} (S_{n_l}^{\beta})^{-1} \\
    & \sum_N \frac{\widetilde{{v}}_{IJ}^{\alpha} \widetilde{{v}}_{IJ}^{\beta}}{\Delta \widetilde{\Omega} - \omega + i\Gamma} + \sum_N \frac{\widetilde{{v}}_{IJ}^{\alpha} \widetilde{{v}}_{IJ}^{\beta}}{\Delta \widetilde{\Omega} + \omega - i\Gamma} \text{,}
\end{split}
\end{equation}
where $c_h$ is the speed of light in the surrounding medium, $Z_h$ is the wave impedance, 
and $k_h$ is the wave number of the host material.\cite{Zerulla.Krstic.ea:Multi-Scale.2022}
Note that these quantities are frequency dependent. $\mathbf{S}$ again denotes the
Cartesian to spherical basis transformation coefficients, and $\Gamma$
is an imaginary broadening factor, preventing divergence
in case $\Delta \widetilde{\Omega} = \omega$.
The index $N$ formally runs over all perturbed excited states, 
which is unfeasible for sizable systems. A direct access of the perturbed excited state 
T-matrix would however require fourth-order response theory.

\section{Computational methods}

To outline the capabilities of calculating excited state
properties using the BSE, we investigate the excited state
absorption spectra of the dipyrrometheneboron difluoride 
(BODIPY) derivative pyrromethene-567 (PM567)
\cite{Costela.Garcia-Moreno.ea:Photophysical.2002}
and the well known bipyridine (bpy)
[Ru(bpy)$_3$]$^{2+}$ complex of ruthenium.
\cite{Yersin.Humbs.ea:Low-lying.1997}
Furthermore, we put the perturbative spin-orbit coupling
approach to test with the demanding mercaptoaryl-oxazoline 
complex [PdI(S-phoz)(IMes)],
\cite{Holzer.Dupe.ea:Mercaptoaryl-Oxazoline.2018} 
for which its spectral features
have been shown to strongly depend on SOC,
\cite{Holzer.Klopper:Ionized.2019} and with the 
[Ir(ppy)$_2$(sip)]$^{+}$ complex featuring a 5d metal center.
\cite{Bi.Yang.ea:Cyclometalated.2020}
The geometries of the metal complexes were optimized using 
the BP86\cite{Becke:Density-functional.1988, Perdew:Density-functional.1986} functional with the def2-SVP basis set.
\cite{Weigend:Accurate.2006}
For the geometry optimizations, the ECP28MWB (Pd, Ru, I)
and ECP60MWB (Ir) effective core potentials were used.
\cite{Andrae.Hauermann.ea:Energy-adjusted.1990}
were used. For PM567, the respective geometries reported in 
Ref.~\citenum{Valiev.Cherepanov.ea:First-principles.2018}
were used.
Subsequent scalar-relativistic (time-dependent) PBE0,
\cite{Perdew.Burke.ea:Generalized.1996,Adamo.Barone:Toward.1999}
$G_0W_0${\makeatletter @}PBE0, 
eigenvalue self-consistent $GW${\makeatletter @}PBE0 
(ev$GW${\makeatletter @}PBE0), and BSE 
calculations were performed using the all-electron x2c-TZVPPall 
basis set.\cite{Pollak.Weigend:Segmented.2017} 
Reference two-component calculations were performed 
using the x2c-TZVPPall-2c basis set.
\cite{Pollak.Weigend:Segmented.2017}
Ground states were tightly converged to errors of less than
10$^{-8}\,$Hartree in energy, and differences in the densities
of less than 10$^{-7}\,$. 
A grid of size 3 was used in DFT calculations.
\cite{Treutler:Entwicklung.1995,Treutler.Ahlrichs:Efficient.1995}
A seminumerical exchange algorithm was used to accelerate
DFT calculations in combination with a very fine grid.
\cite{Holzer:improved.2020}
For the $G_0W_0$ and ev$GW$ calculations, the highest 10/20
occupied and lowest 10/20 unoccupied orbitals (1c)/spinors
(2c) were optimized, and the remaining orbitals/spinors subsequently
shifted. The exact two-component (X2C) transformation was used to 
include relativistic effects.
\cite{Peng.Middendorf.ea:efficient.2013, Franzke.Middendorf.ea:Efficient.2018}
Note that 10 spatial orbitals equals 20 spinors for closed shell
systems. Quasiparticle equations in ev$GW$ were considered converged 
if the change of HOMO and LUMO was lower than 10$^{-5}\,$Hartree.
To evaluate excited state properties, for each molecular
system, the first 16 singlet and triplet excited states
were determined. Subsequently, all excited state transition
moments between these states are extracted to yield 
the excited state absorption spectra.
In case of perturbative spin-orbit coupling, 
the coupling matrix was also constructed from the same 16 singlet
and triplet states in addition to the singlet ground states.
Due to the high numerical effort, 2c BSE 
calculations were carried out using two Nvidia A100 GPUs
to accelerate the construction of matrix-vector products.
\cite{Franzke.Holzer.ea:NMR.2022}
All plotted spectra were obtained by applying a
broadening of 0.05$\,$eV full width at half maximum (FWHM).

All calculations were performed using a development version of 
Turbomole V7.9.\cite{Franzke.Holzer.ea:TURBOMOLE.2023,
Balasubramani.Chen.ea:TURBOMOLE.2020} 

\section{Results and discussion}

\subsection{[Ru(bpy)$_3$]$^{2+}$}
\label{sec:rubpy}

\begin{figure}[htbp]
    \centering
    \includegraphics[width=1.0\columnwidth]{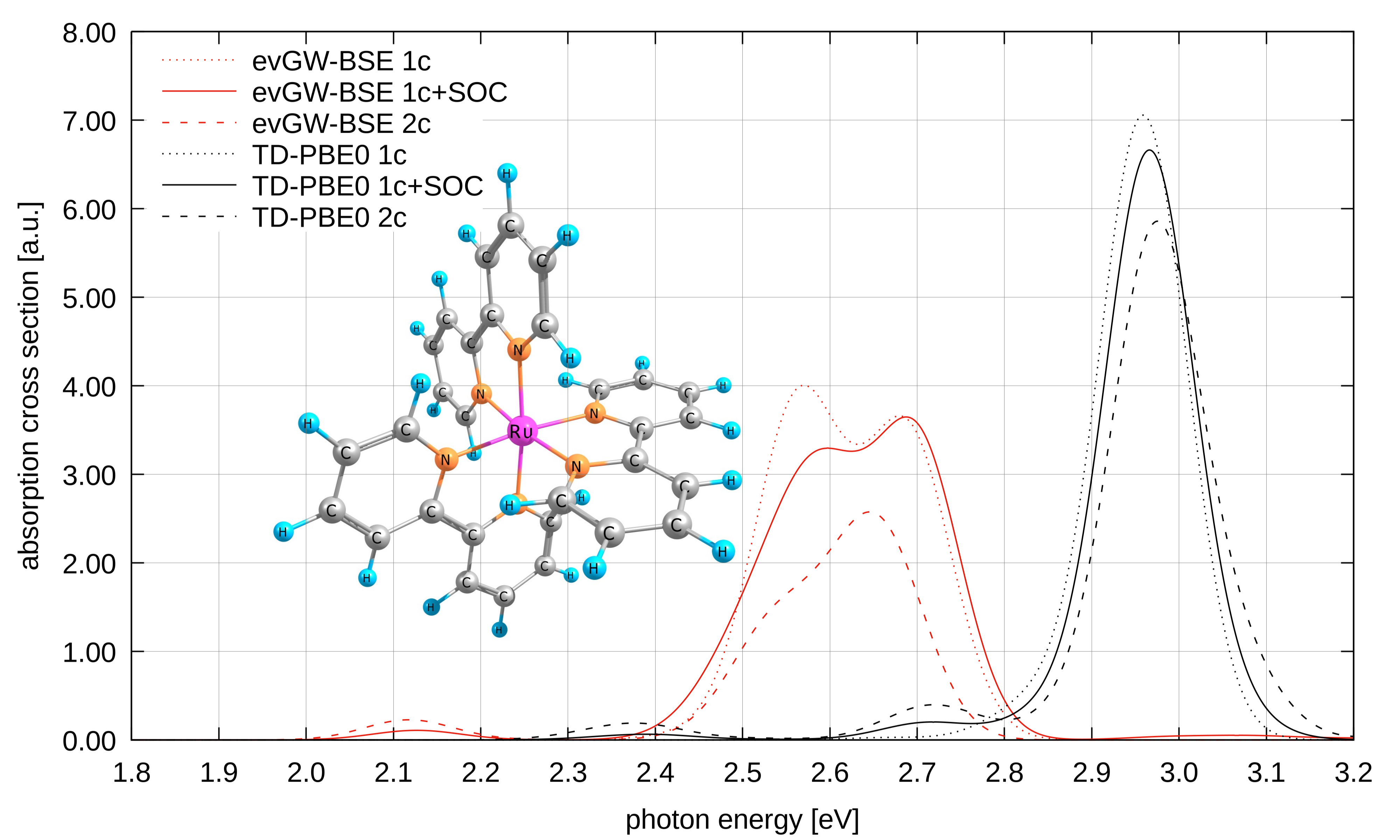}
    \caption{Ground state absorption spectra of [Ru(bpy)$_3$]$^{2+}$
    obtained from scalar relativistic (1c, dotted line), scalar relativistic plus perturbative spin orbit coupling (1c+SOC, solid line)
    and fully relativistic two-component (2c, dashed line) 
    spectra obtained from ev$GW$--BSE and TD-PBE0.}
    \label{fig:Rubpy_GS}
\end{figure}

The tris(bipyridine)ruthenium(II) dication [Ru(bpy)$_3$]$^{2+}$
is one of the most well-known optically active complexes, 
with optical investigations dating back nearly three
decades.\cite{Yersin.Humbs.ea:Low-lying.1997}
The simulated TD-PBE0 and ev$GW$--BSE absorption spectra
are outlined in Fig.~\ref{fig:Rubpy_GS}.
ev$GW$--BSE is in very good agreement with the experimental
measurements of Yersin \textit{et al},
\cite{Yersin.Humbs.ea:Low-lying.1997}
who found the lowest energy band starting at approximately 2.73~eV 
(22,000~cm$^{-1}$). This band is assumed to be a metal-ligand 
charge transfer (MLCT) band, therefore being found at a too low 
energy when TD-PBE0 is used. Both TD-PBE0 and ev$GW$--BSE further 
predict this band to be composed of two excited states, 
with ev$GW$--BSE and TD-PBE0 yielding distinctly different 
oscillator strengths. 
Interestingly, a comparison of the 1c and 2c
methods reveals that these principal absorption bands that define the visible part of the [Ru(bpy)$_3$]$^{2+}$ spectrum are only
moderately influenced by spin-orbit coupling.
At low temperatures, additionally, an emission band at lower
energy is observed. This band is assigned to a triplet
excited state\cite{Yersin.Humbs.ea:Low-lying.1997}
and can therefore only be modelled if spin-orbit coupling is 
included. In the case of [Ru(bpy)$_3$]$^{2+}$, Fig.~\ref{fig:Rubpy_GS}
outlines that our implemented perturbative approach is
in excellent agreement with the full 2c methods. Both excitation energy
and oscillator strength of the triplet band are well reproduced
by our perturbative SOC approach, labelled as 1c+SOC.
The reproduction if the 2c spectrum is achieved slightly better
by the ev$GW$--BSE method than it is for TD-PBE0.

\begin{figure}[htbp]
    \centering
    \includegraphics[width=1.0\columnwidth]{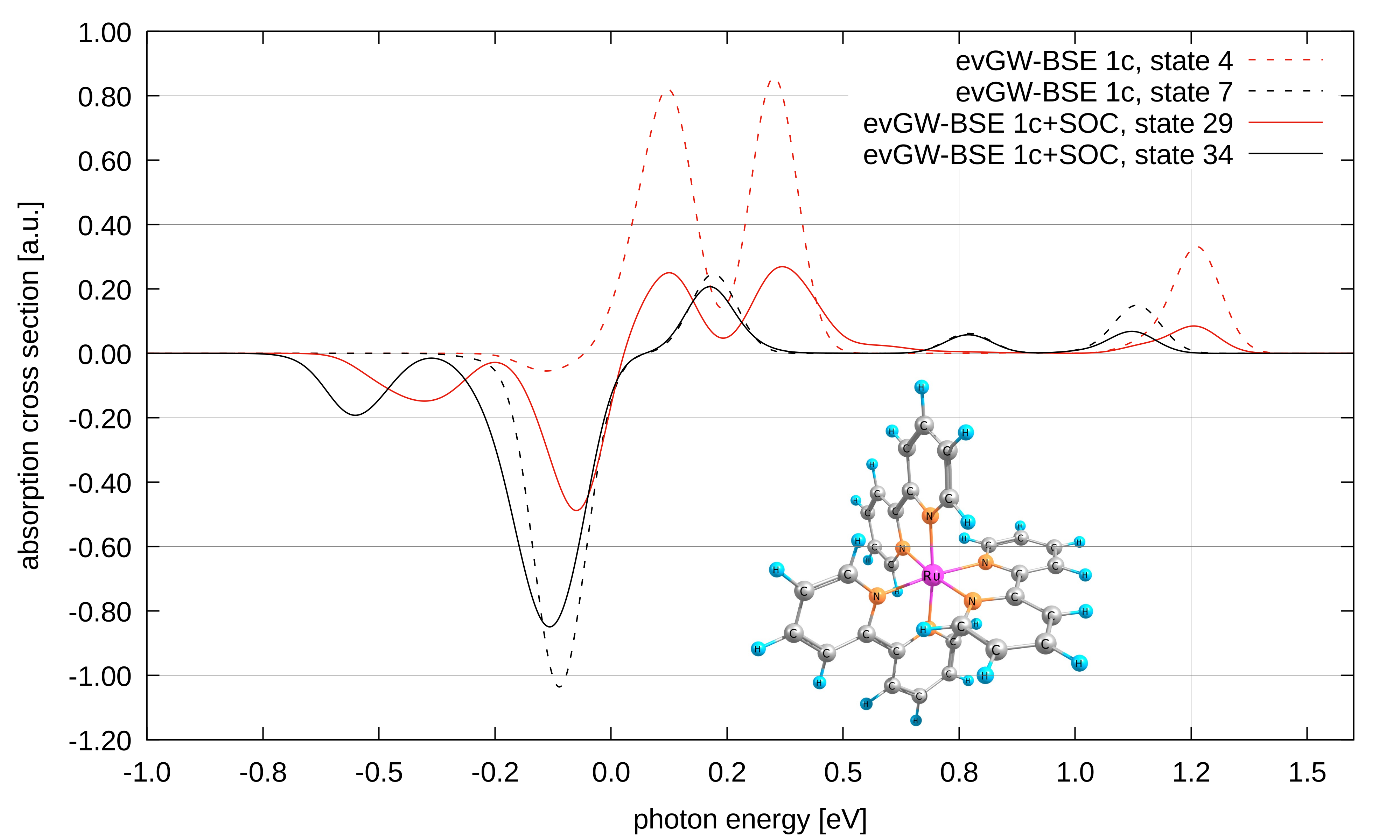}
    \caption{Excited state absorption and emission spectra of 
    [Ru(bpy)$_3$]$^{2+}$
    obtained from scalar relativistic (1c, dotted line) and 
    scalar relativistic plus perturbative spin orbit coupling 
    (1c+SOC, solid line)
    spectra obtained from ev$GW$--BSE. Negative oscillator
    strengths correspond to emission lines.}
    \label{fig:Rubpy_ES_BSE}
\end{figure}

\begin{figure}[htbp]
    \centering
    \includegraphics[width=1.0\columnwidth]{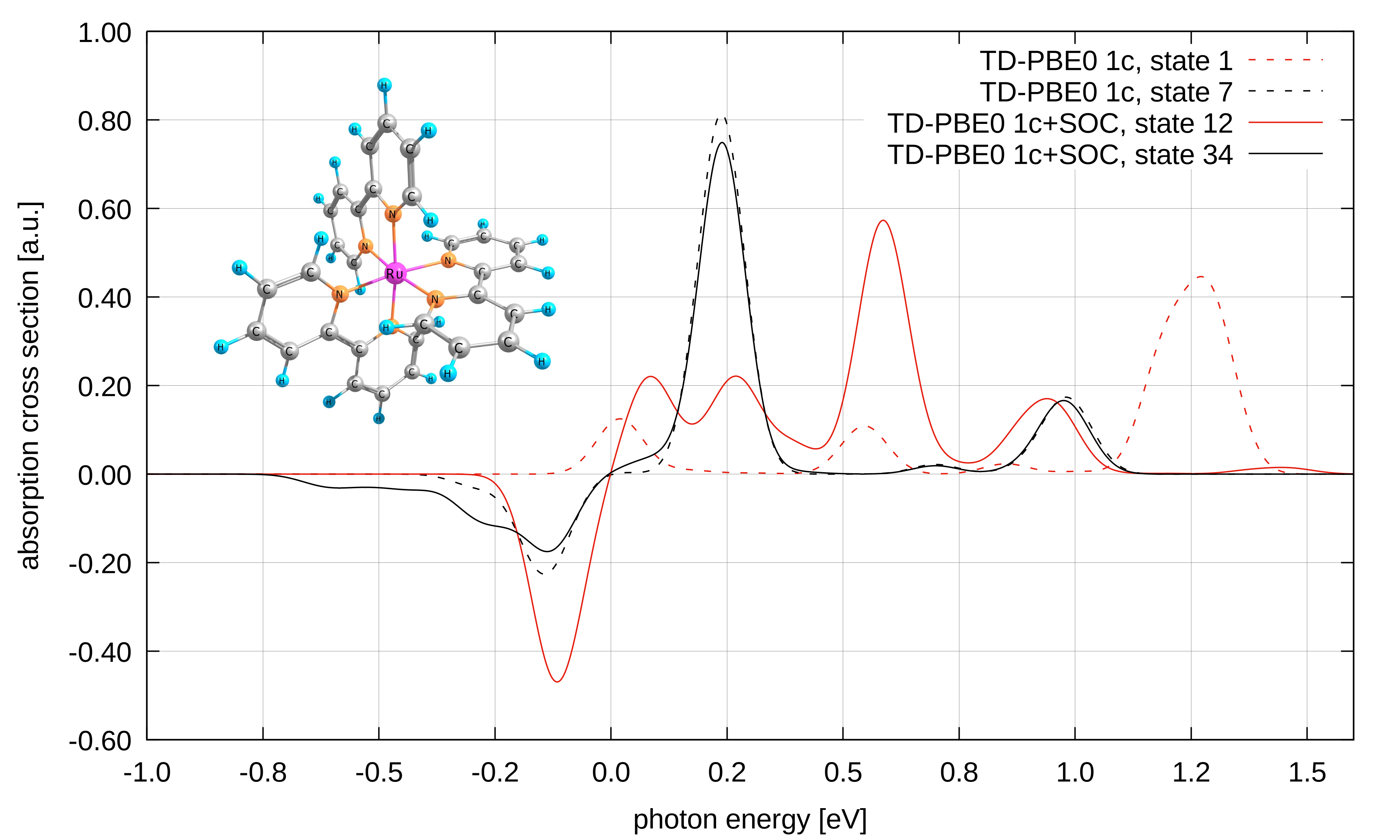}
    \caption{Excited state absorption and emission spectra of [Ru(bpy)$_3$]$^{2+}$
    obtained from scalar relativistic (1c, dotted line) and scalar relativistic plus perturbative spin orbit coupling (1c+SOC, solid line)
    spectra obtained from TD-PBE0. Negative oscillator
    strengths correspond to emission lines.}
    \label{fig:Rubpy_ES_DFT}
\end{figure}

Given the availability and stability of [Ru(bpy)$_3$]$^{2+}$, 
it is one of the few compounds where also experimental 
excited state absorption data is available.
\cite{Milder.Gold.ea:Time-resolved.1988,Hauser.Krausz:excited-state.1987}
Hauser and Krausz found the first excited state absorption
peak for the MLCT excited state at approximately 1.2$\,$eV (10,000$\,$cm$^{-1}$)
\cite{Hauser.Krausz:excited-state.1987}, which is again well
reproduced by ev$GW$--BSE as shown in Fig.~\ref{fig:Rubpy_ES_BSE}.
Additionally, ev$GW$--BSE hints at excited state absorption 
being also important in the infrared region below 0.5$\,$eV, 
though no experimental data is available in this range.
Contrary to the ground state, spin-orbit coupling is 
import to correctly describe the ESA amplitude of the MLCT 
band in this complex, marked using red lines in Fig.~\ref{fig:Rubpy_ES_BSE}. While the peak positions nearly
perfectly align, large deviations in the predicted oscillator
strengths can be found. For TD-PBE0, the ESA spectra of the 
simulations with and without spin-orbit coupling finally bear no
resemblance of each other, underlining the importance of including
SOC when tackling excited state properties of Ru complexes.
Contrary, for a mainly ligand centered band, marked using black
lines in Figs.~\ref{fig:Rubpy_ES_BSE} and \ref{fig:Rubpy_ES_DFT}, 
SOC coupling is only important in the de-excitation regime,
where intersystem crossing can lead to a lower-lying
metal-centered triplet state, which is strictly forbidden 
without spin orbit coupling.

\subsection{PM567}
\label{sec:PM567}

\begin{figure}[htbp]
    \centering
    \includegraphics[width=1.0\columnwidth]{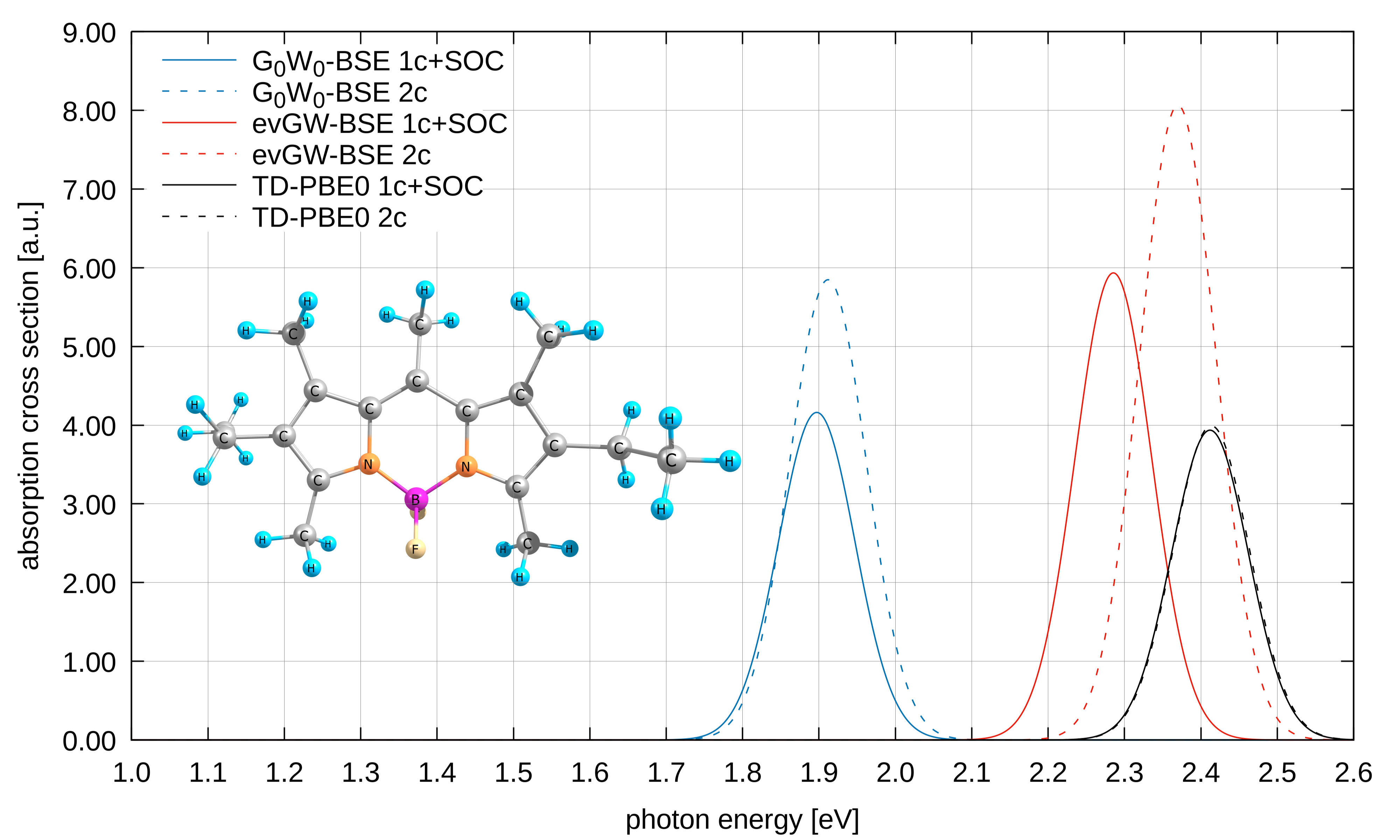}
    \caption{Ground state absorption spectra of the BODIPY derivative PM567
    obtained from scalar relativistic (1c, dotted line), 
    scalar relativistic plus perturbative spin orbit coupling 
    (1c+SOC, solid line) and fully relativistic two-component 
    (2c, dashed line) spectra obtained from $G_0W_0$--BSE, 
    ev$GW$--BSE, and TD-PBE0.}
    \label{fig:pm567_gs}
\end{figure}

The dipyrrometheneboron difluoride (BODIPY) derivatives represent 
another common case that is usually not well described by TD-DFT.
\cite{Momeni.Brown:Why.2015} 
We therefore turn to the PM567 variant, which has recently been
investigated by Valiev \textit{et al}.
\cite{Valiev.Cherepanov.ea:Calculating.2019,
Valiev.Merzlikin.ea:Internal.2023}
They consider Duschinsky, anharmonic, and Herzberg–Teller effects
when describing intersystem crossing rates for PM567. While we will
not dive into these effects in detail, still another important 
detail can be observed in our simulations. Namely, the dependence 
of the BSE spectrum on the underlying $GW$ spectrum.
As shown in Fig.~\ref{fig:pm567_gs}, TD-PBE0 1c and 2c results 
are nearly identical, i.e. spin-orbit effects are not too
important in the absorption spectrum of PM567. However, for $GW$
methods, 1c and 2c results are slightly different, with small deviations
in the peak positions and significant deviations in the 
obtained oscillator strengths. This can be attributed to
differences in the $GW$ step of the calculation. In this step, 
all orbitals enter the construction of the polarizability. 
As BODIPY contains third row elements, the resulting core orbital
energies are significantly influenced by spin-orbit coupling.
While effect is less important for $G_0W_0$, i.e. the HOMO-LUMO gaps
between 1c and 2c $G_0W_0$ differ by less than 0.01$\,$eV, the effect 
is more pronounced for ev$GW$. For the latter, the HOMO-LUMO gap 
is widened by 0.05$\,$eV when full spin-orbit coupling is used 
already in the $GW$ step. This is directly carried over to the 
BSE step, and cannot be corrected by the perturbative SOC correction
as outlined by Fig.~\ref{fig:pm567_gs}. The reason of perturbative 
correction being unable to correct this issue is that it is already
present in the zeroth-order Hamiltonian $\hat{H}^0$ in Eq.~\ref{eq:qdpt},
i.e. the excitation energies. The resulting 1c+SOC results
therefore closely resemble the plain 1c results, similar to TD-PBE0.
Subsequent deviations in the oscillator strengths arising from the 
different screening in $\textbf{W}$\cite{Holzer.Klopper:Ionized.2019} are even 
more pronounced. This effect is especially strong for PM567,
but in line with earlier observations on the dependence
of the BSE oscillator strengths on the underlying $GW$ method.
\cite{Jacquemin.Duchemin.ea:Assessment.2016}
We note that setting the spin-orbit operator to zero in the 2c 
calculation leads to a perfect alignment of the 1c and 2c spectra, 
and we are therefore confident in the numerical validity of the results
shown in Fig.~\ref{fig:pm567_gs}.

\begin{figure}[htbp]
    \centering
    \includegraphics[width=1.0\columnwidth]{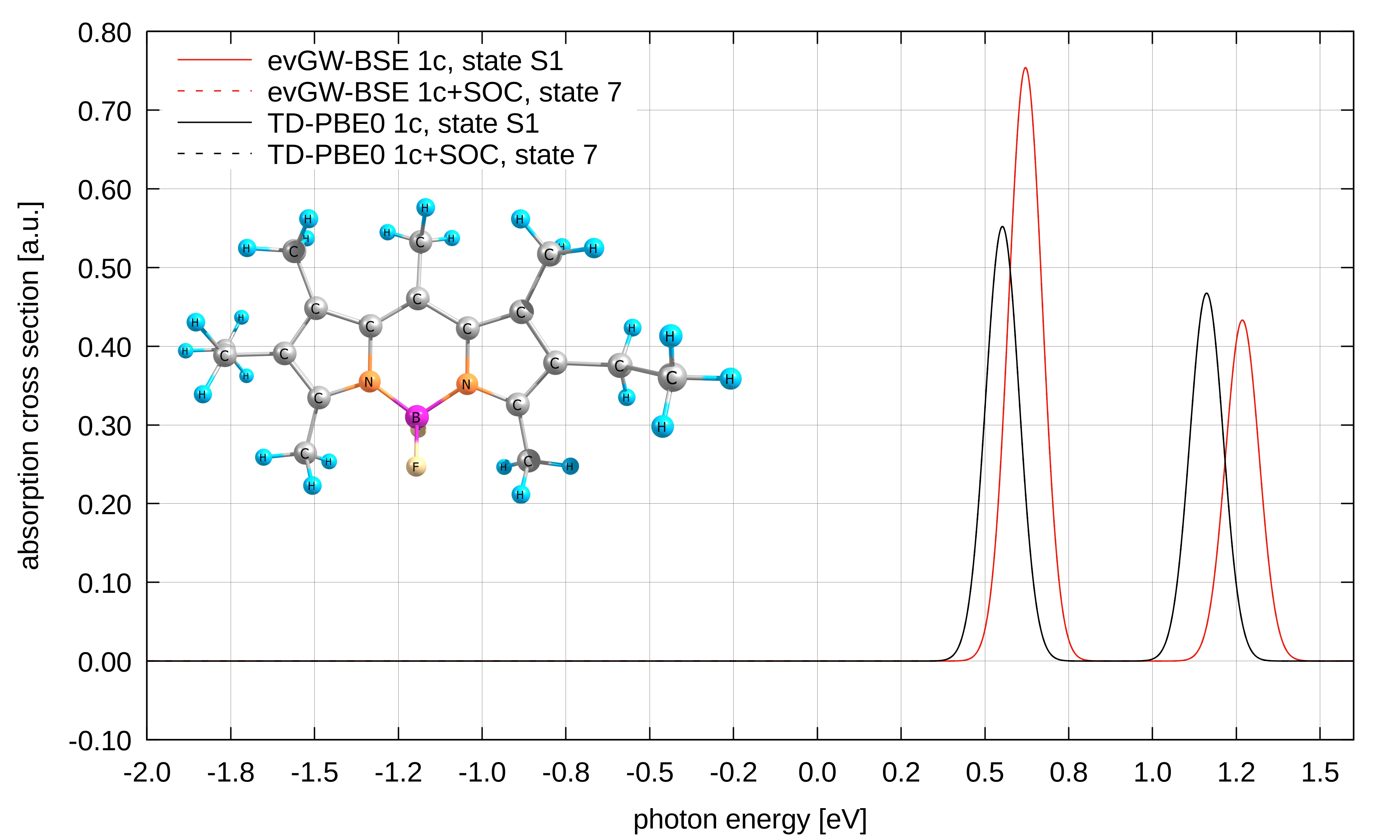}
    \caption{Excited state absorption and emission spectra of the 
    BODIPY derivative PM567 obtained from scalar relativistic 
    (1c, dotted line) and scalar relativistic plus perturbative 
    spin orbit coupling (1c+SOC, solid line)
    spectra obtained from ev$GW$--BSE. Negative oscillator
    strengths correspond to emission lines.}
    \label{fig:pm567_es}
\end{figure}

Fig.~\ref{fig:pm567_es} further outlines that SOC effects can also
not be attributed as the sole or even main reason of the
observed intersystem crossing in PM567. SOC effects on the 0-0
transitions are vanishing, and ISC is therefore dominated by
Franck-Condon and Herzberg–Teller effects as outlined in literature.
\cite{Valiev.Cherepanov.ea:Calculating.2019,
Valiev.Merzlikin.ea:Internal.2023}

\subsection{[PdI(S-phoz)(IMes)]}
\label{sec:pdsphoz}

\begin{figure}[htbp!]
    \centering
    \includegraphics[width=1.0\columnwidth]{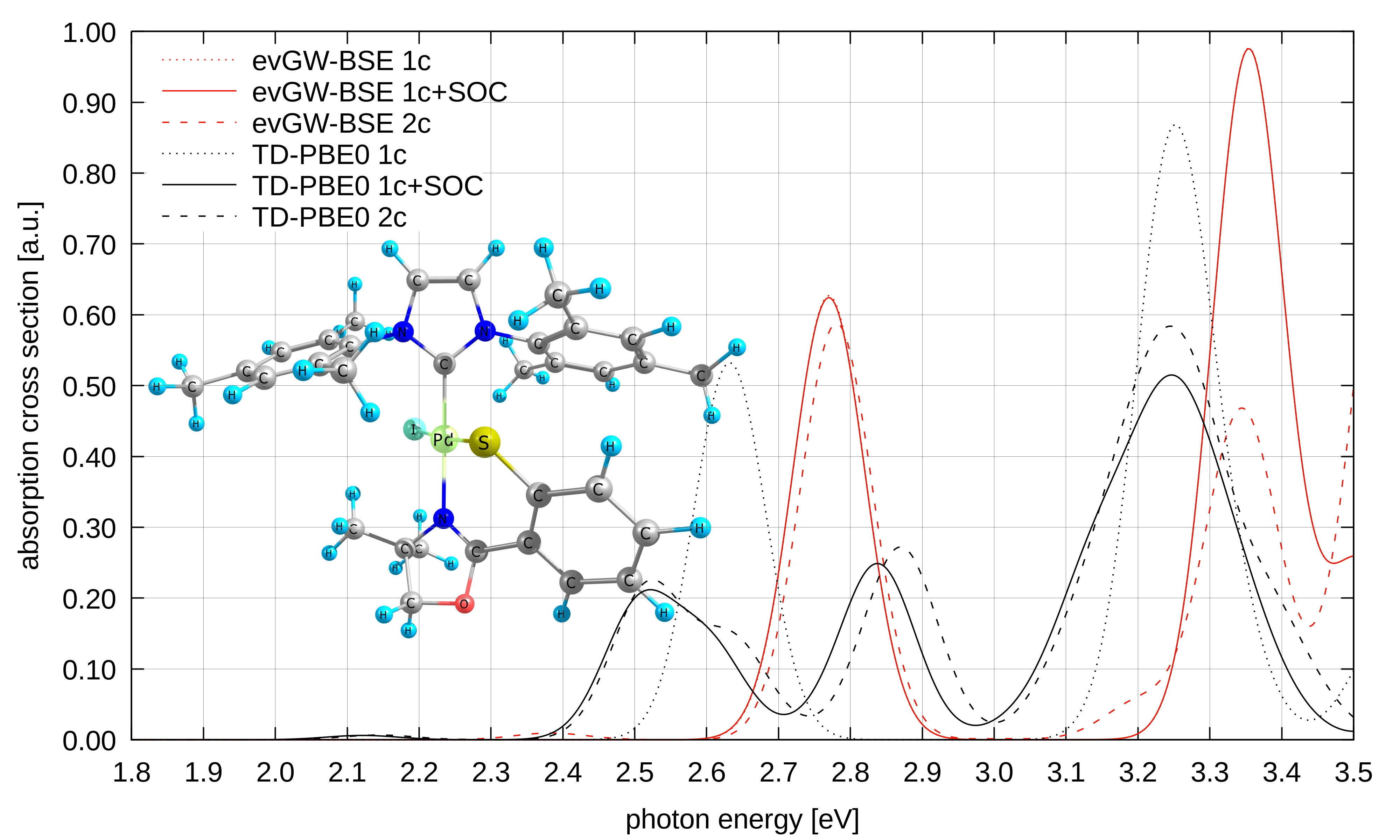}
    \caption{Ground state absorption spectra of [PdI(S-phoz)]
    obtained from scalar relativistic (1c, dotted line), scalar relativistic plus perturbative spin orbit coupling (1c+SOC, solid line)
    and fully relativistic two-component (2c, dashed line) 
    spectra obtained from ev$GW$--BSE and TD-PBE0.}
    \label{fig:pdsphoz}
\end{figure}

The mercaptoaryl-oxazoline complex [PdI(S-phoz)(IMes)] 
\cite{Holzer.Dupe.ea:Mercaptoaryl-Oxazoline.2018}
has been previously investigated by TD-PBE0 and $GW$--BSE,
\cite{Holzer.Klopper:Ionized.2019} and it was concluded
that especially the spectrum of the halogenated iodine 
variant can only be correctly described with SOC being used.
It therefore constitutes an excellent test for the perturbative
SOC approach outlined in Sec.~\ref{subsec:SOCMEs} of this work.
As outlined in Fig.~\ref{fig:pdsphoz}, indeed the effects of
SOC are substantial. The uncorrected 1c spectrum barely has any
resemblance to the full 2c spectrum. For example, for 
uncorrected scalar relativistic TD-PBE0, a single peak is
predicted at 2.63$\,$eV. After turning on SOC, this peak is 
split into two major bands from 2.4 to 3.0$\,$eV with reduced
oscillator strength. Yet, the perturbative SOC procedure
is able to largely correct for this issue. In the predicted 
1c+SOC spectrum, the peak positions are both re-adjusted and the
oscillator strengths accordingly redistributed. This leads to 
an excellent match between the full and perturbative SOC spectra.
[PdI(S-phoz)(IMes)] is a prime example of the importance of SOC
to recover the correct spectrum in cases where SOC is important
and $S_n$-$T_n$ gaps are not too small. While complexes with 
large $S_n$-$T_n$ gaps are usually of no interest in the design of
optical devices due to their low phosphorescence quantum 
yields, they are excellent probes for methodological advancements 
in theoretical applications as outlined by Fig.~\ref{fig:pdsphoz}.

\subsection{[Ir(ppy)$_2$(sip)]$^{+}$}
\label{sec:irppy}

\begin{figure}[htbp!]
    \centering
    \includegraphics[width=1.0\columnwidth]{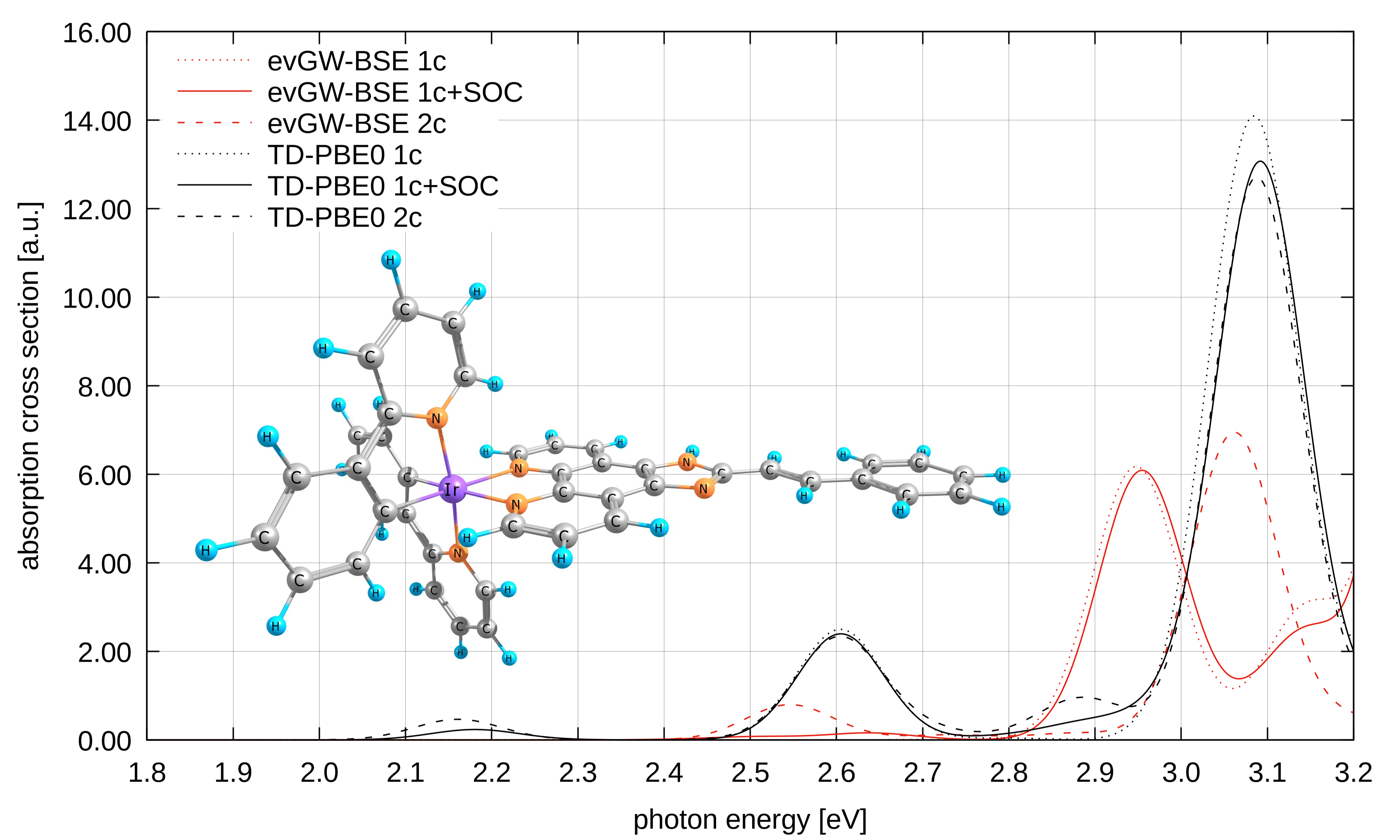}
    \caption{Ground state absorption spectra of [Ir(ppy)]
    obtained from scalar relativistic (1c, dotted line), scalar relativistic plus perturbative spin orbit coupling (1c+SOC, solid line)
    and fully relativistic two-component (2c, dashed line) 
    spectra obtained from ev$GW$--BSE and TD-PBE0.}
    \label{fig:irppy}
\end{figure}

As a final molecular test system, we consider a 
2-phenylpyridine (ppy) complex of Iridium, namely the [Ir(ppy)$_2$(sip)]$^{+}$
molecule.\cite{Bi.Yang.ea:Cyclometalated.2020} This complex
has a notable two-photon absorption cross section, making
it an ideal candidate for further theoretical testing in this direction.
Similar to its parent, the [Ir(ppy)$_3$] complex, it features broad absorption
bands with centers at 2.65$\,$eV (468$\,$nm), 3.07$\,$ (404$\,$nm),
and 3.40$\,$eV (365$\,$nm) that have been assigned
to a series of metal-ligand charge-transfer states.
\cite{Hofbeck.Yersin:Triplet.2010} 
Fig.~\ref{fig:irppy} outlines that including spin-orbit coupling
only slightly alters the overall shape of the absorption spectrum.
Especially TD-PBE0 closely resembles the experimental spectrum
described in Ref.~\citenum{Bi.Yang.ea:Cyclometalated.2020}, 
yielding peaks close to the experimental reference values, 
with only minor differences between 1c, 1c+SOC and full 2c spectra.
Most prominent differences can again be found in the 
lowest energy part, where a transition to the
triplet excited state is strictly forbidden in the 1c calculation, 
while 1c+SOC and 2c yield rather similar finite oscillator strengths.
For ev$GW$, the differences between 1c and 2c are more pronounced, 
which we again likely attribute to changes in the principal gap
stemming from the $GW$ procedure, that is not corrected by the
perturbative SOC step. 

However, Fig.~\ref{fig:irppy} tends to oversimplify the complexity
of the bands. While indeed the shape is rather similar, when SOC
is turned on, many more excited states will contribute to the
band due to state mixing, though the overall intensity 
is not significantly changed, as these changes average out
after broadening. Earlier investigations of the [Ir(ppy)$_3$] 
complex using quasi-degenerate perturbation theory at the 
DFT/MRCI level came to the same conclusion, finding a significant 
amount of state mixing.
\cite{Kleinschmidt.Wullen.ea:Intersystem-crossing.2015}
The last observation is not too surprising given the
5d Ir metal center promotes SOC. Distinct to the [PdI(S-phoz)(IMes)] 
complex discussed in Sec.~\ref{sec:pdsphoz}, the $S_n$-$T_n$ gaps
in [Ir(ppy)$_2$(sip)]$^{+}$ and [Ir(ppy)$_3$] however only
amount to a few 100$\,$cm$^{-1}$. A mixing of these states therefore
barely affects the intensity of the band, leading to the
partially incorrect impression that SOC is not too important
for these iridium complexes.

\section{Conclusion}

We have derived an efficient way to calculate excited state 
properties including perturbative spin-orbit corrections
for the $GW$-Bethe--Salpeter equation method. These advancements
allow for the calculation of excited state absorption
as well as other excited state properties as for example
excited state circular dichroism, optical rotation, or
T-matrices for multiscale modeling.\cite{Zerulla.Krstic.ea:Multi-Scale.2022,Zerulla.Beutel.ea:Multi-Scale.2023,Zerulla.Li.ea:Exploring.} 
Furthermore, using the same route, 
perturbative spin-orbit coupling corrections can be obtained
for both ground and excited state spectra. While we focused on 
the $GW$--BSE method, we also outlined the necessary modifications 
needed to extend perturbative spin-orbit coupling to 
excited state properties within time-dependent DFT.
Given the broad availability of TD-DFT, these adaption can be
useful to enhance existing codes.

To demonstrate the capabilities of our BSE excited state property
implementation, we have investigated four sizable molecular 
systems. First, it could be shown that excited state 
absorption is well described by our $GW$--BSE implementation, 
closely following experimentally observed features in 
[Ru(bpy)$_3$]$^2$. Next, it could be shown that including 
SOC is straightforward for TD-DFT, and leads to 
excellent agreement with fully relativistic 2c TD-DFT calculations.  
For $GW$--BSE, the situation is more complex when perturbative SOC
is to be included. The perturbative SOC treatment only recovers 
state interactions from the BSE part, however does not correct 
for differences entering the zeroth-order Hamiltonian through 
differences in the quasiparticle energies obtained from the respective
1c and 2c $GW$ methods. Further corrections would be needed to fully
align 1c and 2c $GW$--BSE methods, though the differences are usually
well within the errors of obtained excitation energies and 
oscillator strengths. 
\cite{Jacquemin.Duchemin.ea:0-0.2015,Jacquemin.Duchemin.ea:Assessment.2016,Gui.Holzer.ea:Accuracy.2018}
The conclusion of Ref.~\citenum{Jacquemin.Duchemin.ea:Assessment.2016},
stating that a more detailed evaluation of the dependence of
excited state properties on the underlying DFT functional as 
well as the details of the $GW$ method needs to be carried out,
is underlined by our results.

\section*{Supplementary Material}

Standard xyz files with the optimized structures,
as well as machine readable ASCII files 
listing all excited states and oscillator strengths used to
generate Figs.~\ref{fig:Rubpy_GS} to \ref{fig:irppy} 
are contained in the supplementary material.

\section*{Acknowledgements}

C.H.\ gratefully acknowledges the Volkswagen Stiftung for financial support. 

\section*{Author Declarations}

\subsection*{Conflict of Interest}

The authors have no conflicts to disclose.

\section*{Data Availability}

The data that support the findings of this study are available
within the article and its supplementary material.

\section*{References}
%\nocite{*}
\bibliography{literature}% Produces the bibliography via BibTeX.

\end{document}